# New Constraints On Cosmic Polarization Rotation Including SPTpol B-mode Polarization Observations

W.-P. Pan

*Department of of Physics, National Tsing Hua University,
No. 101, Section 2, Kuang-Fu Road, Hsinchu, Taiwan 30013, ROC*
*two.joker@gmail.com*

S. di Serego Alighieri

*Osservatorio Astrofisico di Arcetri, Istituto Nazionale di Astrofisica,
Largo Enrico Fermi 5, I-50125 Firenze. Italia*
*sperello@arcetri.astro.it*

W.-T. Ni

*Department of of Physics, National Tsing Hua University,
No. 101, Section 2, Kuang-Fu Road, Hsinchu, Taiwan 30013, ROC*
*weitou@gmail.com*

L. Xu

*School of Physics and Optoelectronic Engineering, Dalian University of Technology,
No. 2 Linggong Road, Ganjingzi District, Dalian City, Liaoning Province, China, 116024*
*lxxu@dlut.edu.cn*

We present an update of the cosmic polarization rotation (CPR) constraint from the recent SPTpol measurements of sub-degree B-mode polarization in the cosmic microwave background (CMB) of 100 square degrees of sky. Our previous CPR fluctuation constraint from the joint ACTpol-BICEP2-POLARBEAR polarization data is 23.7 mrad (1.36°). With new SPTpol data included, the CPR fluctuation constraint is updated to 17 mrad (1°) with the scalar to tensor ratio $r = -0.05 \pm 0.1$.

## 1. Introduction

A cornerstone of General Relativity (GR) and metric theories of gravity is the Einstein Equivalence Principle[1] (EEP). EEP is an important tool in the studies of new realms of physics such as early Universe, Black Hole Physics and Quantum Gravity. The precision of its empirical validity is crucial in these studies and rests on the possibility of extending also to the EEP the tests of the weak or Galilean equivalence principle[2] (WEP). However, this extension from the WEP to the EEP would not be fully justified if there were a pseudoscalar field coupling to electromagnetism and leading to a violation of the EEP, while obeying the WEP.[3,4] Such pseudoscalar field, if it existed, would also produce a rotation of the polarization angle (PA) independent of frequency for radiation traveling over large distances across the universe,[3,5] the so called cosmic polarization rotation (CPR).

In the premetric formulation of continuous electrodynamics or spacetime electrodynamics,[6] the field strength $F_{kl}$ (***E***, ***B***) and excitation (density with weight +1) $H^{ij}$ (***D***, ***H***) are related by the constitutive tensor density

$$H^{ij} = \chi^{ijkl}(F_{kl}). \tag{1}$$

In general $\chi^{ijkl}(F_{kl})$ is a functional of $F_{kl}$. For local linear medium or for energy much less than Planck energy, (1) is reduced to the following multi-linear (tensor) relation:

$$H^{ij} = \chi^{ijkl} F_{kl}. \tag{2}$$

The Lagrangian for electromagnetic field is

$$L_{\rm I}^{\rm (EM)} = -(1/(16\pi))H^{ij} F_{ij} = -(1/(16\pi))\chi^{ijkl} F_{ij} F_{kl}. \tag{3}$$

$\chi^{ijkl}$ has 21 degrees of freedom with the following permutation symmetry

$$\chi^{ijkl} = -\chi^{jikl} = -\chi^{ijlk}. \tag{4}$$

In the $\chi$-framework (3), from the nonbirefringence of cosmic propagation of electromagnetic wave packet, the constitutive tensor of spacetime for local linear gravity-coupling to electromagnetism must be of core metric ($h^{ik}$) form with an axion (pseudoscalar) $\varphi$ degree of freedom and a dilaton (scalar) $\psi$ degree of freedom:[7]

$$\chi^{ijkl} = \tfrac{1}{2} (-h)^{1/2}[h^{ik} h^{jl} - h^{il} h^{kj}]\psi + \varphi e^{ijkl}. \tag{5}$$

From observations, the nonbirefringence is empirically verified to $10^{-38}$, i.e. to $10^{-4} \times (M_{\rm Higgs}/M_{\rm Planck})^2$.[8] This is significant in constraining the infrared behavior of quantum gravity. Empirically, axion is constrained by the non-observation of cosmic polarization rotation (Sec. 2);[8,9,10] dilation is constrained by the agreement of CMB (Cosmic Microwave Background) spectrum to Planck spectrum.[8,11]

If empirically there was no CPR to high accuracy, the EEP would be verified to high accuracy in this aspect. This would greatly increase our confidence in the EEP and consequently in GR. In fact, CPR could be induced also by violations of other fundamental physical principles (see Ref. [12] for a review), connected with the breaking of symmetry and parity: clearly, if there were any CPR, i.e. if the CPR angle were not zero, it should be either positive for a counter-clockwise rotation, or negative for a clockwise rotation, leading to asymmetry. A number of other theoretical models lead to the existence of this frequency-independent CPR also. They all have the same or equivalent effective Lagrangian and are not easy to be distinguished solely from astrophysical and cosmological polarization observation although different models predict different CPR angles. More precise and comprehensive observations are needed to distinguish models if CPR is detected. Feng et al.[13] proposed CPT violation and dynamical dark energy models. Liu et al.[14] gave constraints on the coupling between the quintessence and the pseudoscalar. Geng et al.[15] proposed a new type of effective interaction in terms of the CPT-even dimension-six term equivalent to what we are discussing to generate CPR, and used the neutrino number asymmetry to induce a nonzero polarization rotation angle in the data of the CMB polarization. They found that in their model, the rotation effect can be of the order of magnitude of 10–100 mrad or smaller. In our original pseudoscalar model, the natural coupling strength $\varphi$ is of order 1. However, the isotropy of our observable universe to $10^{-5}$ may lead to a change of $\Delta\varphi$ over cosmological distance scale $10^{-5}$ smaller. Hence, observations to test and measure $\Delta\varphi$ to $10^{-6}$ will be significant. A positive result may indicate that our patch of inflationary universe has a 'spontaneous polarization' in the fundamental law of electromagnetic propagation influenced by neighboring patches and by a determination of this fundamental physical law we could 'observe' our neighboring patches.

## 2. New CPR constraints including SPTpol B-Mode polarization observations

Within the last scattering region, three processes can produce B-mode polarization or convert E-mode polarization to B-mode polarization in CMB: (i) local quadrupole anisotropies in the CMB due to large scale gravitational waves (GWs);[16] (ii) primordial magnetic field[17] and (iii) CPR due to pseudoscalar-photon interaction. During propagation, three processes can convert E-mode polarization into B-mode polarization: (i)' gravitational lensing,[18] (ii)' Faraday rotation due to magnetic field (including galactic magnetic field); and (iii)' CPR due to pseudoscalar-photon interaction. The cause of both (i) and (i)' is gravitational deflection; the cause of both (ii) and (ii)' is magnetic field. CPR is independent of frequency while Faraday rotation is dependent on frequency. Therefore Faraday rotation can be corrected for, using observations at different frequencies. However, it is negligible at CMB frequencies and corrections do not need to be applied. The deviation (if there is any) of the speed of gravity propagation from the light speed would change the acoustic peak position of CMB B-mode. From the modeling of observed acoustic peak position of CMB B-mode, this deviation could be measured.[19,20] However, the fitting for velocity of gravity propagation is basically orthogonal to the fitting of other parameters (depends more on the amplitude instead of the acoustic peak position) we are going to consider here.[20] In this paper, we shall assume the speed of gravity propagation is the same as that of light propagation. As to the foreground, the Galactic dust B-mode emission needs to be subtracted in the CMB B-mode polarization measurements (e.g., Ref. 21). At present, there are two quantities that can be measured for CPR $\alpha$ – the mean value $|<\alpha>|$ and the fluctuation amplitude $<(\alpha - <\alpha>)^2>^{1/2}$ in a patch of sky or over all sky.

CPR is currently constrained to be less than about a couple of degrees from the UV polarization observations of radio galaxies and the CMB polarization observations -- 0.02 rad for CPR mean value $|<\alpha>|$ and 0.03 rad for the CPR fluctuations $<(\alpha - <\alpha>)^2>^{1/2}$.[8-10] In 2014, we started to look for the CPR fluctuations from the CMB B-mode observations.[22] In our second study,[23] we use the new Planck dust measurement[21] to update our fits of the BICEP2 CPR fluctuation constraint to be 32.8 mrad (1.88°), and the joint ACTpol-BICEP2-POLARBEAR CPR fluctuation constraint to be 23.7 mrad (1.36°).

Recently, Keisler *et al.*[24] have announced a new measurement of the B-mode polarization power spectrum (the BB spectrum) from 100 deg$^2$ of sky observed with SPTpol (a polarization-sensitive receiver installed on the South Pole Telescope) during 2012 and early 2013 including data in spectral bands centered at 95 and 150 GHz. They reported the BB spectrum in five bins in multipole space, spanning the range $300 \leqslant \ell \leqslant 2300$, and for three spectral combinations: 95 GHz × 95 GHz, 95 GHz × 150 GHz, and 150 GHz × 150 GHz. The power spectra are consistent with predictions for the BB spectrum arising from the gravitational lensing of E-mode polarization. The preference for lensed B modes is of 4.9 σ after marginalizing over tensor power and foregrounds.

BB power spectrum from SPTpol,[24] ACTpol,[25] BICEP2/Keck,[26] and POLARBEAR[27] are drawn on Fig. 1 (reproduced from Figure 4 of Ref. 24). The solid gray line shows the expected lensed BB spectrum from the Planck+lensing+WP+highL best-fit model.[28] The dotted line shows the nominal 150 GHz BB power spectrum of Galactic dust emission derived from an analysis of polarized dust emission in the BICEP2/Keck field using Planck data.[26] The dash-dotted line shows the sum of the lensed BB power and dust BB power. From Ref. 21, the contribution of dust emission to the BB power spectrum of SPTpol, ACTpol and POLARBEAR could be neglected at this stage.

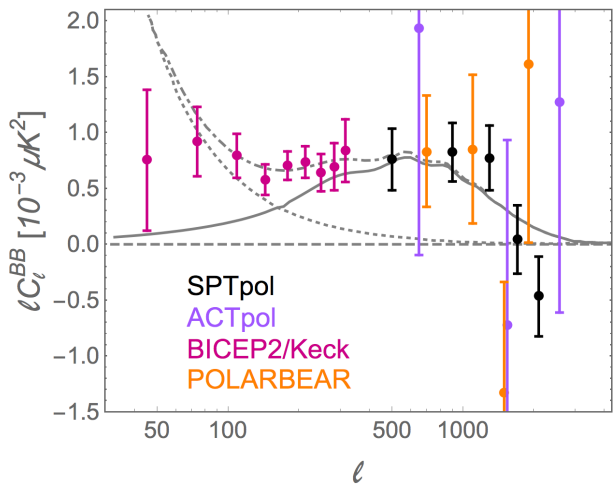

Fig. 1. BB power spectrum from SPTpol, ACTpol, BICEP2/Keck, and POLARBEAR (Ref. 24.)

We fit the B-mode spectrum of various combinations to the two parameters -- the tensor to scalar ratio *r* and the CPR fluctuation power $<(\alpha - <\alpha>)^2>$. The E-mode power spectrum and the lensing B-mode power spectrum are updated from Fig. 3 of Planck 2015 results XIII[29] and Fig. 4 of Planck 2015 results XV[28] respectively. The left column of Figure 2 shows the fitting results of the BICEP2 8 point data; the right column of Figure 2 shows the fitting results of the joint POLARBEAR and BICEP2 8 point data. The left column of Figure 3 shows the fitting results of the joint POLARBEAR, BICEP2 8 point and ACTPOL data; the right column of Figure 2 shows the fitting results of the joint POLARBEAR, BICEP2 8 point, ACTPOL and SPTpol data.

The CPR fluctuation constraint from the joint ACTpol-BICEP2-POLARBEAR polarization data is 23.7 mrad (1.36°).[23] With the new SPTpol data included, the CPR fluctuation constraint is updated to 17 mrad (1°) with scalar to tensor ratio $r = -0.05 \pm 0.1$.

*New Constraints On Cosmic Polarization Rotation Including SPTpol B-mode Polarization Observations*

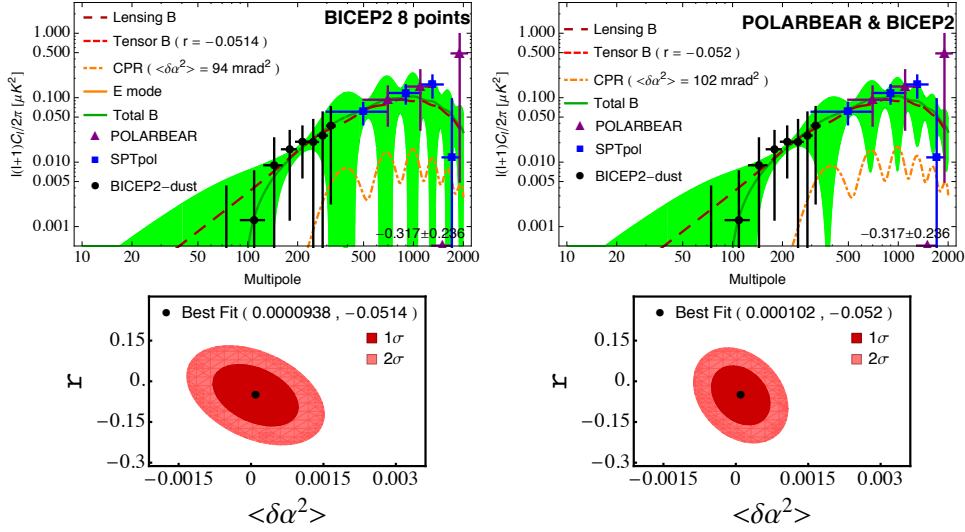

Fig. 2. *Left*: CMB B-Mode power spectrum with data points and models including the tensor-to-scalar ratio $r$, lensing, CPR, and dust contributions. The dust effect is subtracted from the 8 BICEP2 points data by using the PLANCK dust result[x13]. The lowest $l$ BICEP2 data point is not included in the fitting because the PLANCK dust result doesn't cover the the full interval of it. The bottom plot shows the 1σ and 2σ contours on the tensor-to-scalar ratio $r$ and the rms-sum of the CPR angle due to pseudo-scalar photon interaction for the top one. *Right*: The result of the joint POLARBEAR + BICEP2 (8 points) fitting. The value of the third multipole band of POLARBEAR ($l$ from 1300 to 1700) is negative, i.e., $-0.317 \pm 0.236$ μK$^2$ with the value written on the binning interval just above the horizontal axis.

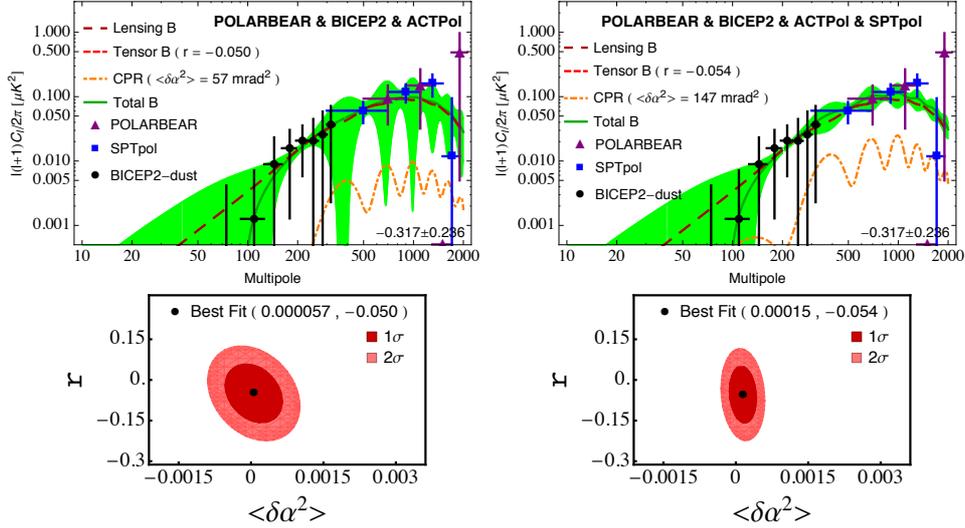

Fig. 3. *Leftt*: The result of the joint POLARBEAR + BICEP2 (8 points) + ACTPol fitting. *Right*: The result of the joint POLARBEAR + BICEP2 (8 points) + ACTPol + SPTpol fitting. From Fig. 2 and Fig. 3, we can see the higher multipole ($l > 400$) experiments (POLARBEAR/ACTPol/SPTpol) are more sensitive to the CPR angle; the BICEP2 ($l<400$) is more sensitive to the tensor-to-scalar ratio $r$.

**Acknoledgements**

We would like to thank Carlo Baccigalupi and Jun-Qing Xia for their helpful discussions and comments on how to extract the dust contribution in specific regions of sky from *Planck* CMB maps and on CPT violating cosmological models respectively.